\documentclass[10pt,journal,twocolumn]{IEEEtran}
\usepackage{graphicx}
\usepackage{amsmath}
\usepackage[noend]{algorithmic}
\usepackage{algorithm}
\usepackage{amsfonts}
\usepackage{amssymb}
\usepackage{bm}
\usepackage{cite}
\usepackage{color}
\usepackage{multirow}
\usepackage{tabularx}
\newtheorem{lem}{Lemma}
\newtheorem{prob}{Problem}
\newcolumntype{L}[1]{>{\raggedright\arraybackslash}p{#1}}
\newcolumntype{C}[1]{>{\centering\arraybackslash}p{#1}}
\newcolumntype{R}[1]{>{\raggedleft\arraybackslash}p{#1}}

\usepackage{times, epsfig}
\usepackage{subfig}
\newlength{\figwidth}
\usepackage{booktabs}

\usepackage{geometry}
\begin{document}
\baselineskip 12 pt

\title{Energy-Efficient Mobile-Edge Computation Offloading over Multiple Fading Blocks}

\author{
Rongfei Fan, Fudong Li, Song Jin, Gongpu Wang, Hai Jiang, and Shaohua Wu
\thanks
{
%
R. Fan and S. Jin are with the School of Information and Electronics, Beijing Institute of Technology, Beijing 100081, P. R. China (email: \{fanrongfei, 15120075709\}@bit.edu.cn). F. Li and H. Jiang are with the Department of Electrical and Computer Engineering, University of Alberta, Edmonton, AB T6G 1H9, Canada (email: \{fudong1,hai1\}@ualberta.ca). G. Wang is with the School of Computer and Information Technology, Beijing Jiaotong University, Beijing 100044, P. R. China (email: gpwang@bjtu.edu.cn). S. Wu is with the Shenzhen Graduate School, Harbin Institute of Technology, Shenzhen, 518055, P. R. China (email: hitwush@hit.edu.cn).
}
\vspace*{-6mm}
}

\maketitle

\begin{abstract}
By allowing a mobile device to offload computation-intensive tasks to a base station, mobile edge computing (MEC) is a promising solution for saving the mobile device's energy. In real applications, the offloading may span multiple fading blocks. In this paper, we investigate energy-efficient offloading over multiple fading blocks with random channel gains. An optimization problem is formulated, which optimizes the amount of data for offloading to minimize the total expected energy consumption of the mobile device. Although the formulated optimization problem is non-convex, we prove that the objective function of the problem is piecewise convex, and accordingly develop an optimal solution for the problem. Numerical results verify the correctness of our findings and the effectiveness of our proposed method.
\end{abstract}


\vspace{-4mm}
\section{Introduction}

Due to the limited computation resources at a mobile device, it is difficult for the mobile device to run computation-intensive applications that require low latency, such as natural language processing, virtual reality (VR), and augmented reality (AR). Mobile edge computing (MEC) provides a solution, in which a mobile device is allowed to offload part of or all its data to a base station for computing, as the base station often has high computation capability. After the base station completes the computing tasks, it feeds back the computed results to the mobile device \cite{Peng}.

Since the offloading is via wireless links, the wireless transmission for offloading involves power consumption and leads to latency. How to minimize power consumption of the mobile device while satisfying latency requirements of computing has been investigated in the literature \cite{Jiandong_Li,Kaibin_Huang,TSIPN,Kaibin_Huang_EH,NTU_2013,Zhengguo_Sheng, Munoz,Xiaowen_Cao,2_Kaibin_Huang}.
For a single mobile device with a single antenna, the work in
\cite{Jiandong_Li} considers that the computation task can be partitioned to two parts, which are computed locally at the mobile device and are offloaded to the base station, respectively. The optimal offloading ratio (i.e., the percentage of the computation task that is offloaded) and the mobile device's transmit power level are derived. Also for a single mobile device, the work in \cite{Munoz} investigates a similar research problem when both the mobile device and the base station have multiple antennas. The work in
\cite{2_Kaibin_Huang} supposes that the base station's CPU alternates between busy state and idle state and assumes that the base station's idle intervals are known in advance.
The offloaded data in each epoch (during which the base station's CPU keeps idle or keeps busy) is optimized so as to minimize the mobile device's total energy consumption.
The work in \cite{Kaibin_Huang} considers multiple single-antenna mobile devices working in time division multiple access (TDMA) mode. Each mobile device's offloading ratio, transmit power allocation, and time ratio for channel access are optimized.
The work in \cite{TSIPN} considers a multiple-input and multiple-output (MIMO) MEC system with one and multiple mobile devices. All the data for computing are offloaded to the base station. Pre-coding matrix is designed for the single-device case, while pre-coding matrix design and the base station's computation resource allocation are investigated for the multiple-device case.
In \cite{Zhengguo_Sheng}, a mobile device can be served by one of multiple base stations. Base station selection is investigated so as to minimize the mobile device's energy consumption.
The work in \cite{Xiaowen_Cao} considers a relay between a mobile device and its base station, and optimizes the offloading ratio and the transmit power levels of the mobile device and the relay.

Note that in all these works \cite{Jiandong_Li,Kaibin_Huang,TSIPN,Zhengguo_Sheng,Munoz,Xiaowen_Cao,2_Kaibin_Huang}, the wireless channel gain is assumed to keep unchanged in the whole process of data offloading.
However, the maximal allowable delay of some applications can be longer than one fading block. For example, the application of VR requires the processing delay to be within 20-60 ms \cite{VR}, while the length of one fading block, within which wireless channel gain can be considered unchanged, can be at the order of 1 ms \cite{TSE}.
Thus, the data offloading process may span multiple fading blocks,
which means that the wireless channel gain varies during the data offloading process, and therefore, the methods in \cite{Jiandong_Li,Kaibin_Huang,TSIPN,Zhengguo_Sheng,Munoz,Xiaowen_Cao,2_Kaibin_Huang} cannot be used.

To the best of our knowledge, only two works in the literature consider data offloading over multiple fading blocks, and both works suppose there is a single mobile device in the system.
The work in \cite{Kaibin_Huang_EH} investigates the problem of optimal offloading over multiple fading blocks for one mobile device that is wireless-powered by the base station. \textcolor{black}
{It is assumed that the channel state information (CSI) of multiple future fading blocks is non-causal, i.e., the CSI of multiple future fading blocks is known in advance, which may not be practical.}
The work in \cite{NTU_2013} also deals with offloading over multiple fading blocks.
Data for computation is not partitioned. Thus, the whole computation task is offloaded to the base station or computed locally, depending on which side leads to less energy consumption.
The wireless channel gain is assumed to take a ``good'' state or a ``bad'' state. The computation time at the base station is not taken into account (i.e., the computation time is assumed to be zero). By approximating the energy consumption for offloading one bit of data as a monomial function, a dynamic optimization problem is formulated to minimize the total energy consumption for offloading, if the mobile device decides to offload rather than locally computing its data.

In this paper, we also investigate the MEC offloading over multiple fading blocks.
{
Different from \cite{Kaibin_Huang_EH}, we assume that the CSI of multiple future fading blocks is causal, i.e., the CSI of multiple future fading blocks is not known in advance.
In addition, the research problem in \cite{Kaibin_Huang_EH} is a convex problem, which is solved by using the KKT condition. But our considered research problem is a non-convex problem.
}
Differences of our work from \cite{NTU_2013} are as follows. 1) We consider that the data for computation can be partitioned and executed in parallel at the mobile device and a base station. This model follows  most of existing works \cite{Jiandong_Li,Kaibin_Huang} and is a more general model compared to the non-partitioning model in \cite{NTU_2013}. 2) We take into account computation time at the base station. 

The contributions of our work are summarized as follows.
Considering computation offloading of a mobile device over multiple fading blocks, we formulate an optimization problem to minimize the total expected energy consumption of the mobile device by deciding the amount of data to be offloaded. The formulated problem is a non-convex optimization problem, and thus, is hard to solve.
{To solve the non-convex problem, we discover and theoretically prove that the objective function of the problem is piecewise convex, and thus, the formulated problem can be optimally solved by first finding solutions for individual intervals within which convexity is kept and then picking up the best solution.}


\section{System Model and Problem Formulation} \label{s:model}

Consider one mobile device with a computation-intensive computation task. The mobile device is connected to a base station, and can offload part of its data  to the base station for computing.
After receiving all the offloaded data, the base station starts computation and then feeds back the computed results to the mobile device.

The wireless channel between the mobile device and the base station is
assumed to be block-faded.
Specifically, the channel gain remains constant in a fading block, but varies randomly and independently from fading block to fading block.
{Denote the {\it normalized channel gain} as  $h$, which is the ratio of the wireless channel gain to the background noise variance.} $h$ is independently and identically distributed over fading blocks with probability density function (PDF) $f(h)$.
Since the maximal allowable latency of the computation task can be much longer than the duration of one fading block,
the computation offloading process may span multiple fading blocks.

{At the beginning of a fading block, the mobile device needs to estimate the normalized channel gain in the fading block (for example, the mobile device sends pilot signals to the base station; then the base station estimates the normalized channel gain and feeds back to the mobile device). The time overhead to estimate the normalized channel gain is normally very short, and thus, is not considered in our work. Nevertheless, it is straightforward to take this time overhead into account in our system model.}

Detailed computation and communication models are given as follows.

\subsection{Computation Model} \label{s:comp}
The computation task, denoted as $\mathcal{T}$, can be described by a profile of three parameters $(T, D, c_0)$, where $T$ is the maximal allowable latency, $D$ is the amount of computation input data (in unit of nat for presentation simplicity), and $c_0$ is the number of CPU cycles required for computing one input nat. With such a description, the total number of CPU cycles required for completing task $\mathcal{T}$ is $c_0 D$.\footnote{
{
As shown in \cite{Kaibin_Huang_EH, NTU_2013, Zhengguo_Sheng}, for a given data size $D$, the number of CPU cycles required for computing the data can be written as $C_{0} D$, where $C_{0}$ is a random variable whose distribution, denoted as $g(C_{0})$, depends on the nature of the application. To facilitate the analysis, a fixed value $c_{0}$ is selected in \cite{Kaibin_Huang_EH, NTU_2013, Zhengguo_Sheng} to guarantee that
$\Pr \left(C_{0} D \leq c_{0} D\right) \geq \left(1 - \varepsilon\right)$
where $\Pr(\cdot)$ means probability and $\varepsilon$ is a predefined small value. Thus, $c_{0}$ can be calculated by using knowledge of $g(C_{0})$. Accordingly, the number of CPU cycles required for computing the data with size $D$ can be expressed as $c_0 D$.
}
}
The computation task $\mathcal{T}$ is separable. Denote the amount of local task (i.e., the data to be computed at the mobile device) and external task (i.e. the data to be computed at the base station) as $D_l$ and $D_e$, respectively. Then we have
\begin{equation} \label{e:sum_data_equation}
D=D_l + D_e.
\end{equation}

For the mobile device,
let $f_l$ denote its computation capability, which is in unit of CPU cycles per second and is also called {\em CPU frequency}.
Similar to \cite{Jiandong_Li}, $f_l$ is adjustable and is not larger than an upper limit denoted as $f_l^U$.
To complete the local task within duration $T$,
the mobile device's computation capability should be set as $f_l=\frac{c_0 D_l}{T}$ (recalling that the number of CPU cycles for computing the local task is $c_0 D_l$).
Due to the constraint $f_l \leq f_l^U$, we have
\begin{equation} \label{e:D_l_upper_bound}
D_l \leq \frac{f_l^U T}{c_0}.
\end{equation}
According to \cite{Jiandong_Li},
the consumed energy for the mobile device to finish the local task with CPU frequency $f_l$ within duration $T$  can be written as
\begin{equation} \label{e:energy_local}
E_l(D_l)=k f_l^3 T= \frac{kc_0^3D_l^3}{T^2}
\end{equation}
where $k$ is a fixed coefficient depending on the architecture of the mobile device's CPU.

At the base station, denote the computation capability for serving the mobile device as $f_e$.
Thus, to complete the external task with $c_0 D_e$ CPU cycles, the associated computational time can be expressed as
\begin{equation}
T_e(D_e)=\frac{c_0 D_e}{f_e}.
\end{equation}
Thus, the offloading should be finished within duration $T-T_e(D_e)$.

\subsection{Communication Model} \label{s:model_comm}
Data offloading is performed over multiple fading blocks.
Denote the time duration of one fading block as $T_f$.
Similar to \cite{Kaibin_Huang}, the time for feeding back the computed results from the base station to the mobile device is negligible and thus, is not considered. This is reasonable since the computed results are usually of small data size and the base station can transmit with high power.
{
The total number of fading blocks for offloading $D_{e}$ data nats is a function of $D_{e}$ and can be written as
}
{
\begin{equation} \label{e:num_blocks_T_f}
N(D_e)= \left\lceil{\frac{T-T_e(D_e)}{T_f}} \right\rceil = \left\lceil{\frac{T f_e- c_0 D_e}{T_f f_e}} \right\rceil
\end{equation}
where $\lceil\cdot\rceil$ is the ceiling function}.
For the ease of discussion,
these \textcolor{black}{$N(D_e)$} fading blocks are indexed inversely, i.e., the mobile device starts offloading data in fading block \textcolor{black}{$N(D_e)$} and ends offloading data in fading block 1. Denote the set of the $N(D_e)$ fading blocks as $\mathcal{N}(D_e)\triangleq \{1, 2, ..., N(D_e)\}$.

Note that the mobile device transmits in partial duration in fading block 1. Denote transmission time duration in fading block $n$ as $t_n(D_e)$. Then we have $t_{n}(D_e)=   T_f$, for \textcolor{black}{$n\in \{ 2,3,...,N(D_e)\}$}, and
\textcolor{black}{
\begin{equation} \label{e:t_n_plus_one}
\begin{array}{lll}
t_{1}(D_e) & = T- T_e(D_e) - (N(D_e)-1)T_f \\
& = \left(T-\frac{c_0 D_e}{f_e}\right) - \left\lceil\left(T- \frac{c_0 D_e}{f_e}\right)/T_f \right\rceil T_f + T_f.
\end{array}
\end{equation}
}

In one fading block, recall that $h$ denotes the normalized channel gain. Consider transmission of $d$ data nats with transmission duration $t$ over the spectrum with bandwidth $W$. By using the Shannon channel capacity formula \cite{TSE},
the transmit power at the mobile device should be
\begin{equation} \label{e:transmit_power}
p(d,h,t,W)=\frac{(e^{\frac{d}{tW}}-1)}{h}
\end{equation}
and the consumed energy can be written as
\begin{equation}
e(d,h,t,W)=p(d,h,t,W)t=\frac{(e^{\frac{d}{tW}}-1)t}{h}.
\end{equation}

The mobile device should offload $D_e$ data nats to the base station within $N(D_e)$ fading blocks. We should minimize the energy consumption in the offloading.
{We have two definitions:
\begin{itemize}
\item $J_{n}(d)$ ($n\in {\cal N}(D_e)$) is defined as the minimum expected energy consumption for transmitting totally $d$ data nats from fading block $n$ to fading block 1;
\item $J_{n}(d, h_n)$ is defined as the minimum expected energy consumption for transmitting totally $d$ data nats from fading block $n$ to fading block 1 given that the normalized channel gain in fading block $n$ is a specific value $h_n$.
\end{itemize}}

{Thus, $J_{n}(d)$ and $J_{n}(d,h_n)$ satisfy the following equation $J_{n}(d)=\int_{0}^{\infty} J_{n}(d,h_{n}) f(h_{n}) dh_{n}$.}

According to Bellman's equation in dynamic programming \cite{Bertskas}, we have the following iterative calculation formulas:
\begin{equation} \label{e:J_with_J_h}
J_{n}(d)=\int_{0}^{\infty} J_{n}(d,h_{n}) f(h_{n}) dh_{n}, ~~n\in {\cal N}(D_e),
\end{equation}
in which
\begin{equation} \label{e:J_n_d_h}
\begin{array}{lllr}
J_{n}(d,h_{n}) = \mathop{\min} \limits_{0\leq d_{n} \leq d}\Big(e(d_{n},h_{n},T_f, W) \\
+ \int_{0}^{\infty} J_{n-1}\left(d-d_{n}, h_{n-1}\right)f(h_{n-1})dh_{n-1} \Big) \\
= \mathop{\min} \limits_{0\leq d_{n} \leq d} \left(e(d_{n},h_{n},t_{n},W)
+  J_{n-1,m}\left(d-d_{n}\right) \right),\\
\quad \quad \quad \quad \quad \quad \quad \quad \quad \quad \quad \quad \quad \quad \quad  \quad \quad \quad
\forall n>1.
\end{array}
\end{equation}
and $J_{1}(d, h_{1})= e(d,h_{1},t_{1}, W).$

In (\ref{e:J_n_d_h}), $d_n$ means the offloaded data nats within fading block $n$.
Then by following the iterative calculation formulas in (\ref{e:J_with_J_h}) and (\ref{e:J_n_d_h}),  $J_{N(D_e)}(D_e)$ can be calculated, which is the minimum expected energy consumption for transmitting totally $D_e$ data nats from fading block $N(D_e)$ to fading block 1. To achieve $J_{N(D_e)}(D_e)$, in fading block $n$ ($n\in \mathcal{N}(D_e)$), with $h_n$ measured and $d$ remaining data nats to be offloaded, the mobile device only needs to determine $d_n$ by solving the optimization problem defined in (\ref{e:J_n_d_h}).

\subsection{Problem Formulation}
Given the system model, the total energy consumption of the mobile device to complete the task $\mathcal{T}$ is
\textcolor{black}{
$\left(J_{N(D_e)}(D_e) + E_l(D_l)\right)$}.
Our target is to minimize the total energy consumption by setting $D_l$ and $D_e$ optimally before the mobile device starts the data offloading, i.e., to solve the following optimization problem
\begin{prob} \label{p:original}
\begin{align}
\mathop{\min}\limits_{D_e, D_l} \quad  & \textcolor{black}{J_{N(D_e)}(D_e) + E_l(D_l)} \nonumber \\
\text{s.t.} \quad & D_e \geq 0, \\
            &  D_l \geq 0, \\
& \text{Constraints}~(\ref{e:sum_data_equation}), (\ref{e:D_l_upper_bound}). \nonumber
\end{align}
\end{prob}


\section{Optimal Solution}
Problem \ref{p:original} is not a convex optimization problem since $N(D_e)$ in \textcolor{black}{$J_{N(D_e)}(D_e)$ involves the ceiling} function of $D_e$.
In this section, we will show how to solve Problem \ref{p:original} optimally.
The following lemmas can be expected.

\begin{lem} \label{lem:J_convex}
With $d\in [0, D_{e}]$, $J_{n}(d)$ and $J_{n}(d,h_{n})$ are convex with $d$ for $n\in \{1,2,...,N(D_{e}) -1\}$.
\end{lem}
\begin{IEEEproof}
We use the induction method.
For $n=1$, it can be easily checked that
\begin{equation} \label{e:J_1_h}
\begin{array}{ll}
J_{1}(d,h_{1}) & =e(d,h_{1},t_{1}, W) \\
&=\frac{t_{1}(D_{e})}{h_{1}}\left(e^{\frac{d}{t_{1}(D_{e})W}} - 1\right)
\end{array}
\end{equation}
and
\begin{equation} \label{e:J_1}
\begin{array}{ll}
& J_{1}(d) \\
=& \int_{0}^{\infty} J_{1}(d,h_{1}) f(h_{1})d h_{1} \\
=& t_{1}(D_{e}) \sigma^2 W \left(e^{\frac{d}{t_{1}(D_{e}) W}} - 1 \right) \int_{0}^{\infty} \frac{1}{h_{1}} f(h_{1})d h_{1},
\end{array}
\end{equation}
both of which are convex functions with $d$.

Now,
suppose both $J_{n}(d)$ and $J_{n}(d,h_{n})$
are convex with $d$ for $n\in  \{1,2,...,N(D_{e}) -2\}$. Then it can be derived that
$J_{n+1}(d,h_{n})$ is still convex with $d$ since the infimal convolution of convex functions is still convex \cite[Theorem 5.4]{Rockafellar}.

Given that $J_{n}(d,h_{n})$ is convex with $d$, it is straightforward to see the convexity of $J_{n}(d)$ with $d$ for $n\in \{1,2,...,N(D_{e}) -1\}$ since $J_{n}(d)$ can be interpreted as nonnegative weighted sum of $J_{n}(d,h_{n})$ with weighting coefficient $f(h_{n})dh_{n}$, which is still a convex function \cite{S_Boyd}.

This completes the proof.
\end{IEEEproof}

\begin{lem} \label{lem:J_convex_n_plus_one}
$J_{N(D_{e})}(D_{e})$ is piecewise convex with $D_{e}$.
\end{lem}
\begin{IEEEproof}
From (\ref{e:t_n_plus_one}), we notice that $t_{1}(D_{e})$ is piecewise linear decreasing function with $D_{e}$. Suppose
$\mathcal{I} \triangleq \{0, 1, ..., I^*\}$ where $I^*$ is the maximal integer $i$ such that $(T- iT_f)$ is larger than 0. Define
\begin{equation*}
\mathcal{D}_i\triangleq
\left\{\begin{array}{lrlr}
\left(\left(T-(i+1)T_f\right)\frac{f_e}{c_{0}}, \left(T-iT_f\right)\frac{f_e}{c_{0}}\right], \\
\quad \quad \quad \quad \quad \quad \quad \quad \quad \quad \quad
\forall i\in \mathcal{I} \setminus \{I^*\}, \\
\left(0,  \left(T-iT_f\right)\frac{f_e}{c_{0}}\right], i=I^*.
\end{array}\right.
\end{equation*}
Then when $D_{e}$ falls within $\mathcal{D}_i$, we have
\begin{equation}
t_{1}(D_e)
=T - iT_f - \frac{c_{0} D_{e}}{f_e}.
\end{equation}


Without loss of generality, we assume $D_{e}$ falls into $\mathcal{D}_{i}$, where $i \in \mathcal{I}$. Thus $N(D_{e})=i+1$.
In this case, denote the optimal solution of $d$ achieving $J_{i+1}(D_{e}, h_{i+1})$ as $d_{i+1}^*(D_{e},h_{i+1})$.
Suppose $D_e^{\dag} \in \mathcal{D}_{i}$ and $D_e^{\ddag} \in \mathcal{D}_{i}$, thus $i+1=N(D_e^{\dag})=N(D_e^{\ddag})$.
For any $\alpha \in [0,1]$,
the convexity of $J_{i+1}(D_{e}, h_{i+1})$ can be proved in (\ref{e:convexity_proof}) on the top of the next page,
\begin{figure*}
\begin{equation} \label{e:convexity_proof}
\begin{array}{ll}
\alpha J_{i+1}\left(D_e^{\dag}, h_{i+1}\right) + \left(1-\alpha\right) J_{i+1}\left(D_e^{\ddag}, h_{i+1}\right) \\
= \alpha \cdot e\left(d_{i+1}^*(D_e^{\dag},h_{i+1}), h_{i+1}, T_f, W\right)
+ (1-\alpha) \cdot e\left(d_{i+1}^*(D_e^{\ddag},h_{i+1}), h_{i+1}, T_f, W\right) \\
\quad \quad + \alpha  J_{i}\left(D_e^{\dag}-d_{i+1}^*(D_e^{\dag},h_{i+1})\right)
+ \left(1-\alpha\right) J_{i}\left(D_e^{\ddag}-d_{i+1}^*(D_e^{\ddag},h_{i+1})\right)\\
\overset{\text{(a)}}{\geq} e\left(\alpha d_{i+1}^*(D_e^{\dag},h_{i+1}) + (1-\alpha)d_{i+1}^*(D_e^{\ddag},h_{i+1}), h_{i+1}, T_f , W\right) \\
\quad \quad +  J_{i}\left(\alpha \left(D_e^{\dag} - d_{{i}+1}^*(D_e^{\dag},h_{{i}+1})\right)
+ \left(1-\alpha\right) \left(D_e^{\ddag} - d_{{i}+1}^*(D_e^{\ddag},h_{{i}+1})\right)\right) \\
\overset{\text{(b)}}{\geq} e\left( d_{i+1}^*(\alpha D_e^{\dag} + (1-\alpha)D_e^{\ddag},h_{i+1}), h_{i+1}, T_f , W\right)
 + J_{i}\left(\alpha D_e^{\dag} + (1-\alpha) D_e^{\ddag}-d_{i+1}^*(\alpha D_e^{\dag} + (1-\alpha)D_e^{\ddag},h_{{i}+1})\right) \\
= J_{{i}+1}\left(\alpha D_e^{\dag} + \left(1-\alpha\right)D_e^{\ddag}, h_{{i}+1}\right).
\end{array}
\end{equation}
\hrule
\end{figure*}
where (a) is due to the convexity of $J_{i}(d, h_{i})$ with $d$ according to Lemma \ref{lem:J_convex} and the convexity of $e(d,h,t,W)$ with $d$, and (b) holds since
\begin{equation*}
\begin{array}{ll}
 & \left(\alpha d_{i+1}^*(D_e^{\dag},h_{i+1}) + \left(1-\alpha\right) d_{i+1}^*(D_e^{\ddag},h_{i+1}) \right) \\
 & \in [0, \alpha D_e^{\dag} + (1-\alpha) D_e^{\ddag}]
\end{array}
\end{equation*}
and $d_{i+1}^*(\alpha D_e^{\dag} + (1-\alpha)D_e^{\ddag},h_{i+1})$ is the optimal solution of $d_{i+1}$ for achieving $J_{i+1}(\alpha D_e^{\dag} + \left(1-\alpha\right)D_e^{\ddag}, h_{i+1})$.

This proves the convexity of $J_{i+1}(D_{e},h_{i+1})$ with $D_{e}$ for $D_{e} \in \mathcal{D}_{i}$.
Thus $J_{N(D_{e})}(D_{e},h_{N(D_{e})})$ is a piecewise convex function with $D_{e}$.

Then by following the similar discussion in the proof of Lemma \ref{lem:J_convex}, the piecewise convexity of $J_{N(D_{e})}(D_{e})$ with $D_{e}$ can be also proved.

This completes the proof.
%
%
%
\end{IEEEproof}

According to Lemma \ref{lem:J_convex_n_plus_one} and by checking (\ref{e:energy_local}), the objective function of Problem \ref{p:original} is also a piecewise convex function with $D_e$.
Thus to solve Problem \ref{p:original}, we only need to solve one convex optimization problem (whose global optimal solution is achievable) in each interval of $D_e$, i.e., for $D_e \in \mathcal{D}_i$, $\forall i \in \mathcal{I}$, and select the minimum total energy consumption among $|\mathcal{I}|$ intervals.
Specifically, define the following problem:
\begin{prob} \label{p:reformulate}
\begin{align}
E^*(i) \triangleq \mathop{\min}\limits_{D_e} \quad  & \left( J_{i+1}(D_e) + E_l(D-D_e) \right) \nonumber \\
\text{s.t.} \quad & \left(D - \frac{f_l^U T}{c_0}\right) \leq  D_e \leq D, \\
            &  D_e \in \mathcal{D}_i.
\end{align}
\end{prob}

To find $E^*(i)$, bisection search method can be used since both $J_{i+1}(D_e)$ and $E_l(D-D_e)$ are convex with $D_e$ and Problem \ref{p:reformulate} is a convex optimization problem.

Then we only need to select the minimum $E^*(i)$ among $i\in \mathcal{I}$, which is the minimal expected energy consumption of the mobile device. In other words, Problem \ref{p:original} is solved optimally.

\section{Numerical Results} \label{s:numerical_results}


In this section, numerical results are presented.
Similar to \cite{Jiandong_Li,Kaibin_Huang, TSIPN}, system parameters are as follows:
$c_0=40$, $T=20$ms, $T_f=2$ms,
$D=4 \times 10^4$ nats,
$f_l^U=0.5$GHz, $f_e=1$GHz, $k=10^{-23}$, and $W=1$MHz.
The normalized channel gain $h$ is exponentially distributed (i.e., Rayleigh fading is assumed) with mean value 100.

\begin{figure}
\begin{center}
\includegraphics[angle=0,width=0.33 \textwidth]{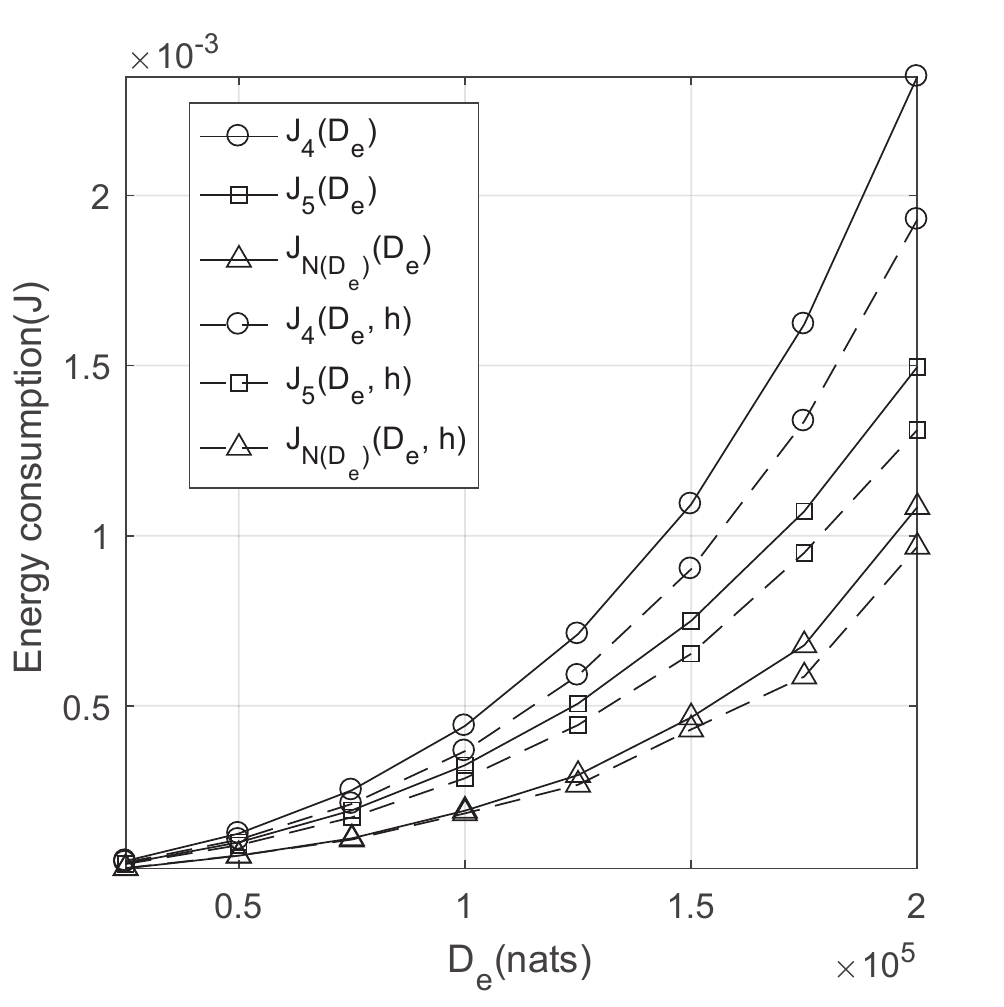}
\end{center}
\caption{Verification of Lemma \ref{lem:J_convex} and Lemma \ref{lem:J_convex_n_plus_one}.}
\label{f:verify_lemma}
\end{figure}

In Fig. \ref{f:verify_lemma}, the functions $J_{N(D_e)}(D_{e})$, $J_{N(D_e)}(D_{e}, h)$, $J_{n}(D_{e})$, and $J_{n}(D_{e},h)$ are plotted for $n \in \{4, 5\}$
and $h=100$.
It can be seen that
the functions $J_{n}(D_{e})$ and $J_{n}(D_{e},h)$ are convex with $D_{e}$, the functions $J_{N(D_e)}(D_{e})$ and $J_{N(D_e)}(D_{e},h)$ are piecewise convex with $D_{e}$,
which support the claims in Lemma \ref{lem:J_convex} and Lemma \ref{lem:J_convex_n_plus_one}, respectively.

\begin{figure}
\begin{center}
\includegraphics[angle=0,width=0.33 \textwidth]{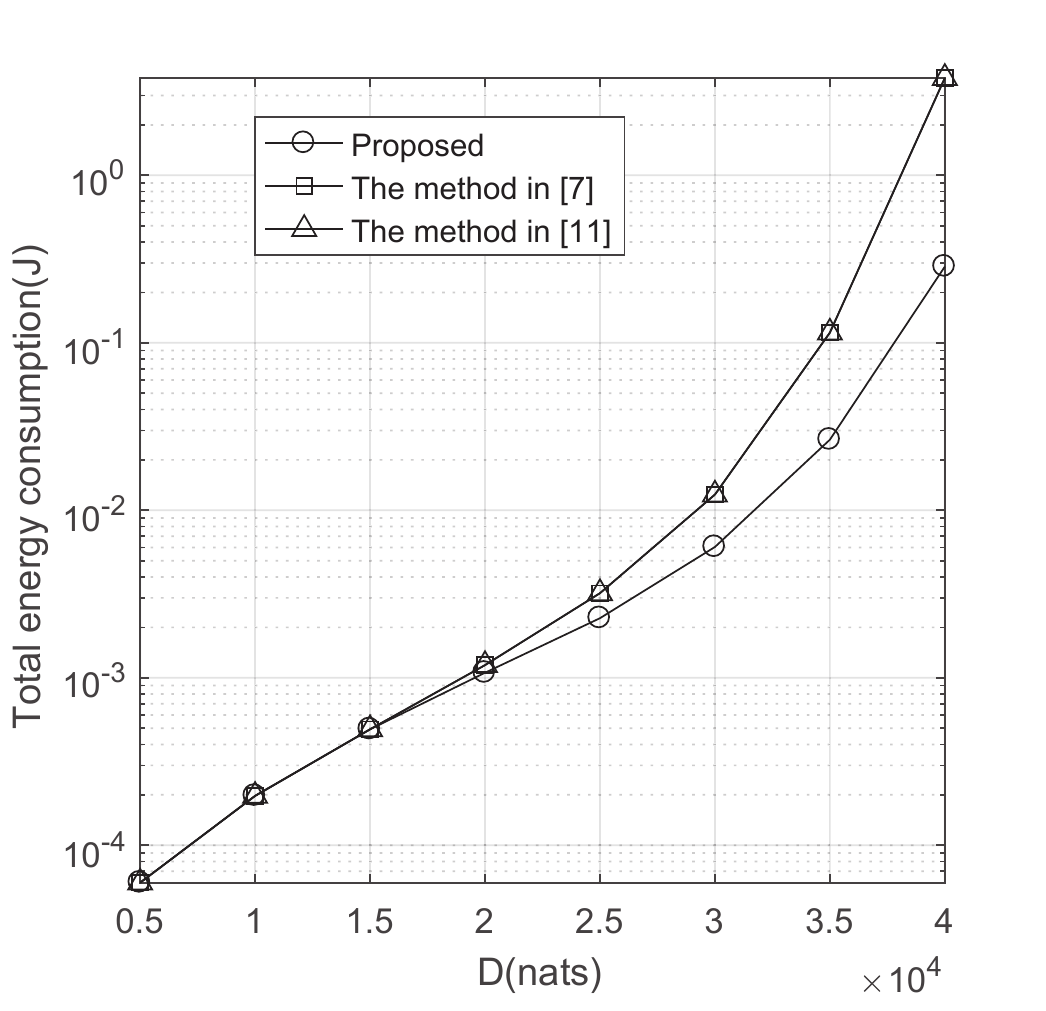}
\end{center}
\caption{Total energy consumption versus $D$.}
\label{f:single_vs_D}
\end{figure}

\begin{figure}
\begin{center}
\includegraphics[angle=0,width=0.33 \textwidth]{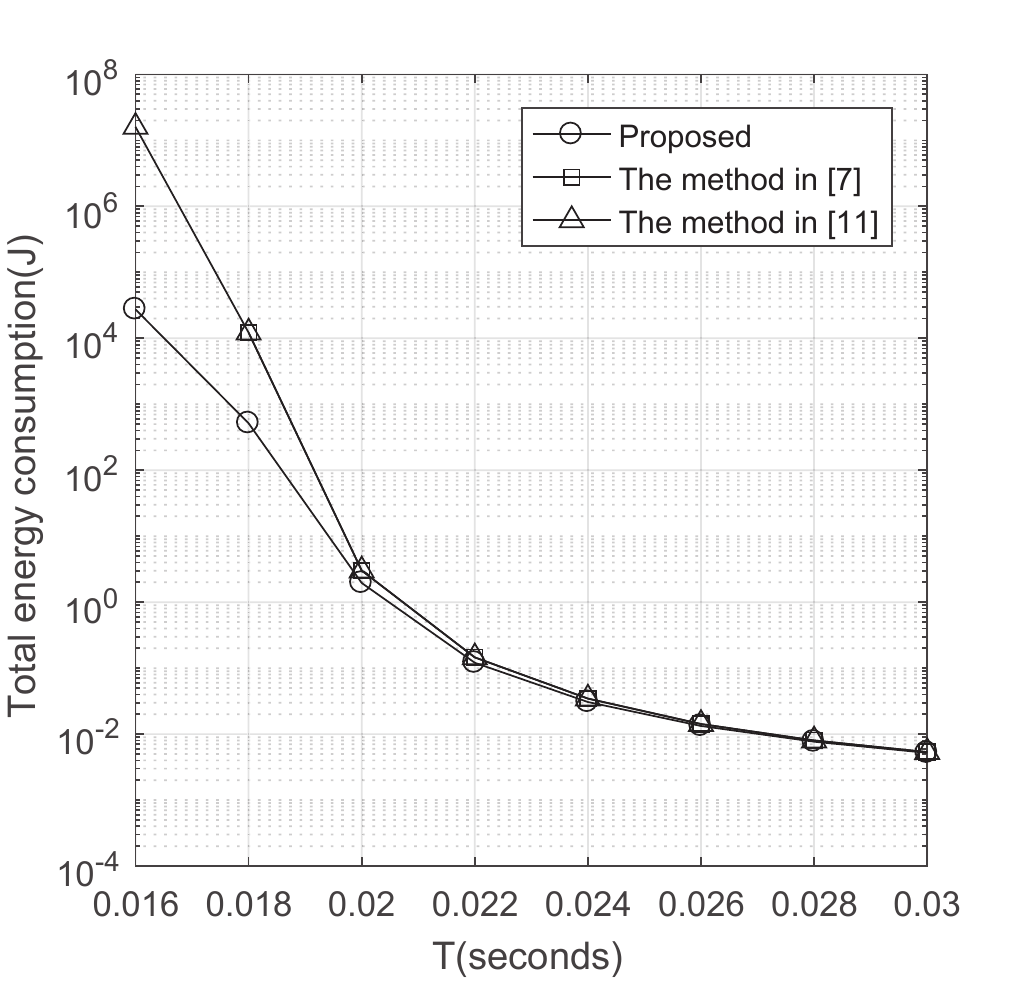}
\end{center}
\caption{Total energy consumption versus $T$.}
\label{f:single_vs_T}
\end{figure}

\begin{figure}
\begin{center}
\includegraphics[angle=0,width=0.32 \textwidth]{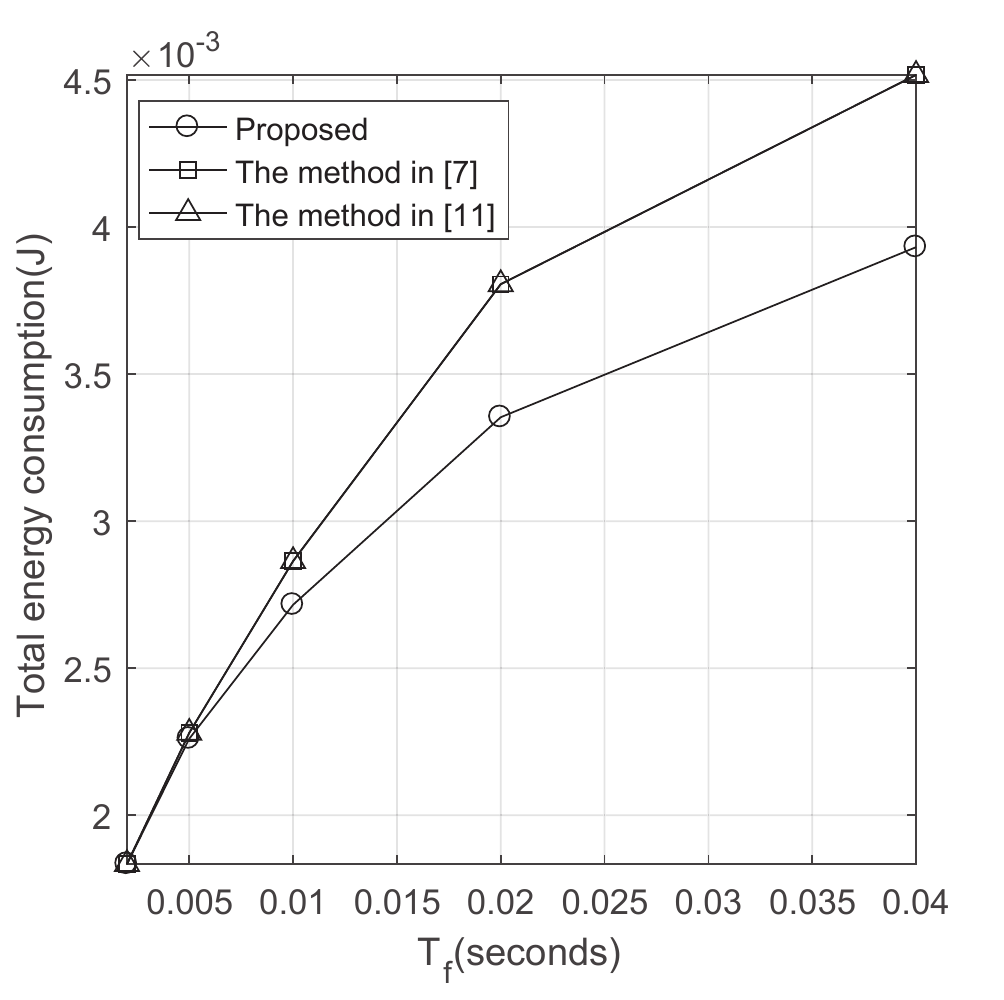}
\end{center}
\caption{Total energy consumption versus $T_f$.}
\label{f:single_vs_Tf}
\end{figure}

In Fig. \ref{f:single_vs_D}, the total energy consumption achieved by our proposed method is 
plotted for $D\in [5\times 10^3, 4\times 10^4]$.
Fig.~\ref{f:single_vs_D} also shows the total energy consumption of the method in \cite{TSIPN}, which offloads all the data to the base station, and the method in \cite{NTU_2013}, which computes all data locally or offloads all data to the base station, whichever has less energy consumption.
It can be found that as $D$ (the amount of data to be processed)  goes higher, the total energy consumption grows, which is intuitive.

{
In Fig. \ref{f:single_vs_T}, the total energy consumption achieved by our proposed method and the methods in \cite{TSIPN,NTU_2013} are plotted versus the deadline $T$.
It can be seen that as $T$ increases, the total energy consumption tends to decrease. This is due to the fact that as the deadline $T$ for processing data  becomes less urgent, the mobile device can have more time to complete its local computing and its data offloading, both of which can help to save energy consumption.
}

{
In Fig. \ref{f:single_vs_Tf}, $T$ is set as 40ms.
The total energy consumption achieved by our proposed method and the methods in \cite{TSIPN,NTU_2013} are plotted versus $T_f$.
It shows that as the duration of one fading block $T_f$ decreases, the total energy consumption decreases.
This can be explained as follows.
When there are more fading blocks within duration $T$,
the mobile user has more
flexibility in adjusting the transmit power over time, i.e., transmit with high power in case of good channel condition and transmit with low power in case of bad channel condition. Thus, more efficient utilization of the transmit power is achieved, which leads to energy saving.
}

{
From Fig. \ref{f:single_vs_D}, Fig. \ref{f:single_vs_T}, and Fig. \ref{f:single_vs_Tf},
it can be seen that our proposed method always has less energy consumption than the methods in \cite{TSIPN,NTU_2013}.
The reason is as follows.
The methods in \cite{TSIPN,NTU_2013} set $D_{e}=D$ or $D_{l}=D$, which means that they only provide a feasible solution to Problem \ref{p:original}. Our method provides the optimal solution to Problem \ref{p:original}, and thus, has superior performance than the methods in \cite{TSIPN,NTU_2013}.
}

\begin{figure}
\begin{center}
\includegraphics[angle=0,width=0.33 \textwidth]{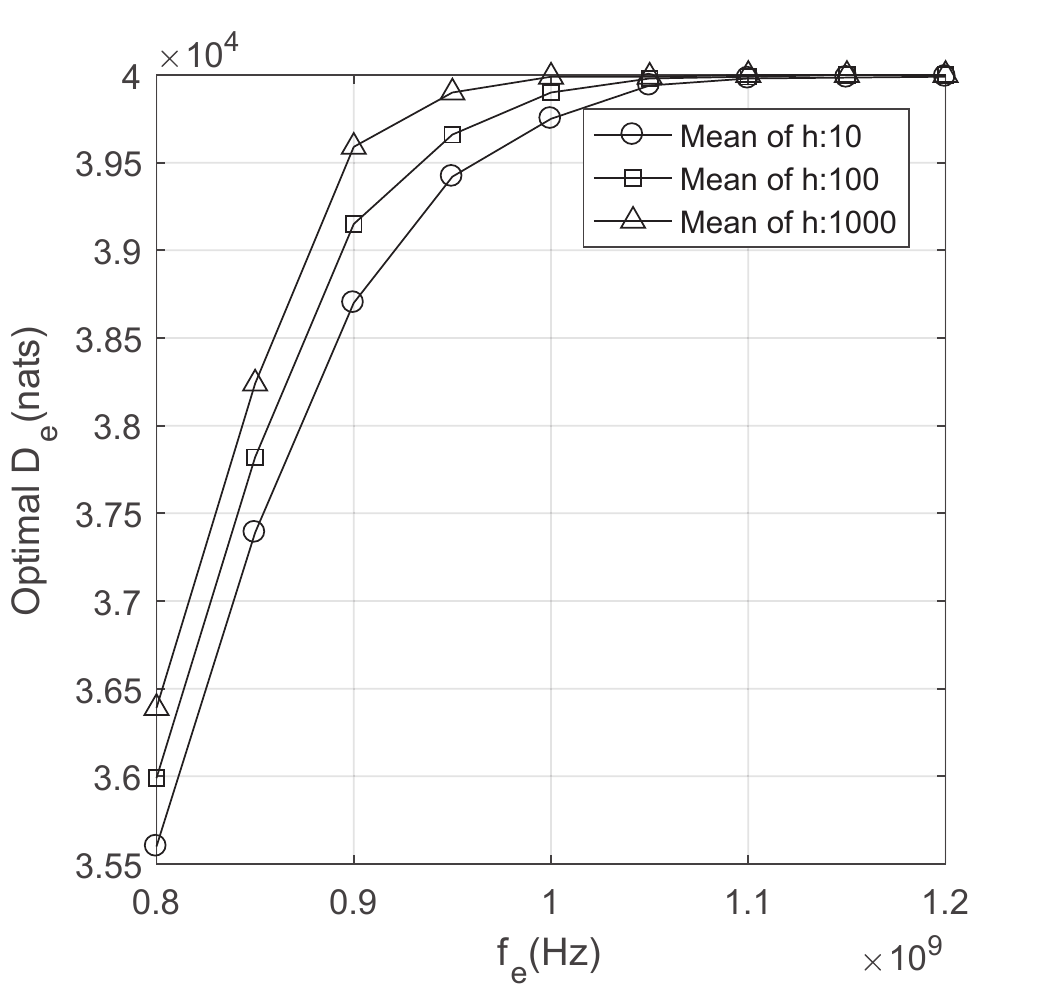}
\end{center}
\caption{Optimal offloading data $D_{e}$ versus $f_e$.}
\label{f:single_optimal_solution}
\end{figure}

{
Fig. \ref{f:single_optimal_solution} plots the optimal offloading data $D_{e}$ under various computation capability $f_e$ and various mean values of normalized channel gain $h$.
From Fig. \ref{f:single_optimal_solution}, it can be seen that as the computation capability $f_e$ goes up, the mobile device tends to offload more data to the base station until all the data is offloaded to the base station.
It can be also seen that as the mean of $h$ grows, the mobile device would like to offload more data to the base station. Indeed, with better wireless channel, the mobile device can send more data to the base station without increasing its power consumption.
}


\section{Conclusion} \label{s:conclusion}
In this paper, we study the energy-efficient offloading in MEC over multiple fading blocks.
An offloading strategy is proposed, which targets minimizing the total expected energy consumption of the mobile device, by selecting the amount of data nats for offloading.
An optimization problem is formulated, which is non-convex.
We show that the objective function of the optimization problem is piecewise convex and develop an optimal solution for the problem.
Since privacy is an important issue for the application of MEC \cite{Tu_1,Tu_2}, in the future, by combining some interesting discussions on security and privacy over data offloading as in \cite{Tu_3,Tu_4,Tu_5},
we wish to extend our work in MEC.


\ifCLASSOPTIONcaptionsoff
  \newpage
\fi

\end{document}